\documentstyle[prd,aps]{revtex}

\begin{document}
\draft

\title{How much energy do closed timelike curves in $2+1$ spacetimes need?} 
\author {Manuel H. Tiglio\thanks{Electronic address: tiglio@fis.uncor.edu}}
\address{Facultad de Matem\'{a}tica, Astronom\'{\i}a y F\'{\i}sica,
 Universidad Nacional de C\'{o}rdoba. Ciudad Universitaria. 5000 
C\'{o}rdoba, Argentina.}
\maketitle

\begin{abstract}
By noticing that, in open $2+1$ gravity, polarized surfaces cannot converge 
in the presence of timelike total energy momentum
 (except for a rotation of $2\pi$),  we give a simple argument which shows that, 
quite generally, closed timelike curves cannot
exist in the presence of such energy condition.
\end{abstract}


\section{Energy conditions and polarized surfaces}
There exist different  types of causality violations in General Relativity (GR).  
One of them corresponds to spacetimes such as Goedel's universe, where there
are closed timelike curves (CTC) passing through each point of spacetime. The
causality violation set is not a result of the evolution of certain initial
data, but rather it exists ``since ever''. There is certain evidence,
provided by the fact that ``we are not being invaded by hords of tourists
coming from the future'', that our universe is not of this kind.

Nevertheless, GR allows for causality violations that ``do not exist since
ever'', but, instead, are generated through spacetime evolution. 
In these cases there exists a Cauchy Horizon ${\cal H}$ (we shall always refer 
to, say, future Cauchy Horizons; the case of
past Cauchy Horizons is, of course, identical), that can be compactly generated or not. 
${\cal H}$ is said to be compactly
generated (CGCH) if its generators, when directed to the past, always enter a compact region 
and remain there forever. A spacetime
with a CGCH is a possible characterization of time machines for the following two reasons:
\begin{itemize}
\item If an otherwise causally well behaved spacetime is changed in a compact 
region such that a Cauchy Horizon ${\cal H}$ appears as a result,
then ${\cal H}$ is compactly generated \cite{haw}. 
\item  Conversely, a CGCH violates strong causality \cite{kay}.
\end{itemize}

For obvious reasons, it is interesting to know under what conditions can ${\cal H}$ 
be non empty. 
It can be seen that the weak energy condition (WEC) must be violated in an open 
spacetime with a CGCH \cite{haw} . In this sense, 
the construction of this kind of time machines needs ``quantum matter'' 
(or the simultaneous creation of a singularity). There is 
a large amount of semiclassical work in this direction, which we do not intend 
to review here. 

Spacetimes with non compactly generated Cauchy Horizons are allowed by classical GR 
(as opposed to compactly generated ones), but it is not clear under which
conditions they should or should not exist.

We can make progress along these lines working in symmetric models,  such that 
we can reduce the problem to one in $2+1$ gravity. Thus, in what follows we
shall restrict ourselves to these low dimensional  models, moreover to open
ones, i.e. with non compact (and simply connected) spatial sections (typically,
${\cal R}^2$). A possible definition of energy momentum (EM) in these
spacetimes is via holonomies. In this way, the total EM is timelike, spacelike
or null, according to whether parallel transport of vectors around loops that
enclose all the matter is defined by a rotation, a boost, or a null rotation,
respectively. The following results can be obtained in $2+1$ \cite{chi}:
\begin{itemize}
\item Under quite general conditions, a CGCH not only
violates strong causality, but also stable causality, since there exist at
least one closed null geodesic (this does not necessarily occur in 3+1, as
emphasized in reference \cite{chr}).
\item Even if one allows for WEC
violations, under certain conditions on the relationship between positive and
negative masses, a CGCH cannot exist if the total EM is timelike (except
when it is a rotation of $2\pi$).
\end{itemize}

The aim of this paper is to give a
result similar to the last one above mentioned, but for positive masses and
non compactly generated horizons. Namely: the original calculations of Gott
\cite{got} showed that in the  spacetime of two particles that gravitationally
scatter each other, certain inequality that involve the masses and
velocities of the particles 
is sufficient for the existence of CTCs. That this inequality is also a
necessary condition can be seen from Cutler's analyisis on the global
structure of these spacetimes \cite{cut}. This inequality, in turn, can be
reexpressed as spacelike total EM , and, in summary, the spacetime of two
particles does not have CTCs if the EM is not spacelike. Kabat \cite{kab} has
further analyzed systems with more particles, and conjectured that as a
general property, CTCs cannot exist in the absence of spacelike EM. Menotti
and Seminara \cite{men} have given a proof of this conjecture for systems 
with rotational symmetry, but, unfortunately, this assumption does not hold
either in solutions like Gott's one or in other ones with different
 number of
particles. Headrick and Gott \cite{hea} have also shown a result related to
Kabat's conjecture: if a CTC is deformable to infinity, then its holonomy
cannot be timelike, except for a rotation of $2\pi$. 

Below we give an
argument which shows that, quite generally, this conjecture is true.
Basically, the argument is the following: if a Cauchy Horizon exists, it can
be obtained as limit of polarized surfaces; on the other hand, these surfaces
cannot converge if the total EM is timelike (except when it is a rotation of
$2\pi$), leading in this way to a contradiction.

The rest of this paper is
devoted to a more detailed description of this simple idea, and heavily relies
on the works of Cutler \cite{cut} and of Carroll et al \cite{car}, to which
the reader can refer for further details. Also, some of the tools here used
are of the kind of those used in \cite{chi}, but in that reference the
exposition is somewhat more detailed. Along this work we implicitly use some
basic properties of curves, CTCs, and causality that can be seen in, e.g.,
\cite{wal} or \cite{haw2}. Finally, a comprehensive review of CTCs in $2+1$
can be found in \cite{hea}. 

The notion of polarized surfaces was originally
introduced by Kim and Thorne in their analysis of vacuum fluctuations and 
whormholes \cite{kim}, and it is widely used in works that study the
stability/instability of Cauchy Horizons under quantum test fields.

The n-th polarized surface $\Sigma (n)$ is defined as the set of points through
which passes a selfintersecting null geodesic (SNG), i.e. a null geodesic
that returns to the same point of spacetime, but possibly with a different
tangent vector, that circles $n$ times the system (this is explicited below).
Its utility as a ``Cauchy Horizon finder'' is a consequence of the following
property:
\begin{equation}
\lim_{n \rightarrow \infty} \Sigma(n) = {\cal H}  \label{pol} 
\end{equation}
Cutler has used this criterion to obtain the
global structure of Gott's spacetime, and it has survived a non trivial check
of self consistency, since Cutler finds that the region where there are CTCs
disappears if the total EM is not spacelike, a fact that is known from other
reasons (e.g., a time function can be globally defined).

For simplicity let
us start discussing the case of two particles, the generalization will be
straightforward. Let us suppose that there are CTCs in this spacetime,
restricted to a region delimited by a Cauchy Horizon ${\cal H}$. It is easy
to see that the CTCs must circle both particles. Thus, the CTCs can be
characterized by the number of times they encircle them, the winding number
$n$. The same holds for the SNGs, and the $n$-th polarized surface $\Sigma
(n)$ is, thus, defined as the set of points through which passes a SNG with
winding number $n$.

We first choose a point $q \in {\cal H}$ and a curve
$\gamma$ which starts at $q$ and ends at some point $p_1$, and  is completely
contained in the region which contains CTCs (except for $q$, which is not in
the region of CTCs but, rather, in its boundary) . That is, 
$\gamma : [0,1]  \rightarrow {\cal M}$, 
with ${\cal M}$ the spacetime
manifold, such that $\gamma (0)=q$ and $\gamma (1)=p_1$. Since $p_1$ is in
the region where there are CTCs, there exists a CTC ${\cal C}_1$ that passes
through $p_1$; this CTC circles, say, $n$ times the pair of particles. We now
 approach $p_1$ to $q$ along $\gamma$, while smoothly deforming the entire
curve ${\cal C}_1$, keeping $n$ fixed, At a certain point, this deformation
will no longer be possible, and the curve that we were deforming will result
in a SNG ${\cal G}_{n}$ that starts and ends at a point $q_1 \in \Sigma(n)$
(it is not clear when will this procedure converge to a closed curve, but
if it does, one can see that it must converge to a SNG). We now take ${\cal
C}_1$ and we move along it twice, obtaining a curve ${\cal C}_{2}$ that passes
through $p_2 (=p_1)$. Repeating the whole procedure, we obtain ${\cal
G}_{2n}$ and a point $q_2 \in \Sigma (2n)$ that is closer to ${\cal H}$, i.e.
there exists a neighborhood ${\cal O}$ of $q$, such that $q_2 \in {\cal O}$
but $q_1 \notin {\cal O}$.  Thus, a point $q_n \in  \Sigma (n)$ will be
closer (in the topological sense just mentioned) to ${\cal H}$ than another
one $q_m \in \Sigma(m)$ with $m<n$. Thus, the sucession $\left\{ q_n \right\}$
converges to $q$, and, in this way, one expects that (\ref{pol}) holds. 

So we find that as a necessary condition for the existence of CTCs, the
polarized surfaces should converge.

In the process $\Sigma (n) \rightarrow {\cal
H}$, the initial tangent to the SNG ${\cal G}_n$, $k^{(i)}_n$, and the final
one, $k^{(f)}_n$, must approach the tangent to the horizon, $k$ (which is a
null vector, since ${\cal H}$ is a null hypersurface). That is, $k^{(f)}_n
\rightarrow k$ and $k^{(i)}_n \rightarrow k$. Since ${\cal G}_n$ is a
geodesic, its tangent is parallel transported, i.e., $k^{(f)}_n = A
k^{(i)}_n$, with $A \in {\cal SO} (2,1)$. The crucial point is that
$A={\cal L}^{n}$, with ${\cal L}$ the holonomic operator that defines the total
EM. Now,  $k$ must be a fixed null direction, i.e. a null eigenvector of
${\cal L}$. So ${\cal L}$ must have at least one null eigenvector. It is easy
to see that if ${\cal L}$ is spacelike or null, it has two and one null
eigenvectors, respectively; and if ${\cal L}$ is timelike it has no null
eigenvector, except when it is the identity (which must correspond to a
rotation of $2 \pi$,  because for a rotation of angle zero the spacetime
would be the well behaved vacuum flat metric). Thus, if the total
EM is timelike and it is not the identity, there cannot be any fixed null
directions and we have reached a contradiction and arrived at our main result.

For more general situations, e.g., if there are an arbitrary
number of particles, one must first recall that every subsystem has timelike
EM if the total EM is timelike \cite{car}. With this property in hand, one 
can then repeat the whole construction and show that the polarized surfaces
cannot converge if the total EM is timelike.

The property that the
evolution of data with timelike total EM is free
 of singularities and/or
Cauchy horizons seems to be a general feature that is not even restricted to
particle-like solutions, but that, instead, also holds for fields coupled to
gravity (the simplest case of this statemente being Einstein-Rosen waves, or a
massless scalar field coupled to gravity, if seen as a $2+1$ system). A
rigorous proof for  vacuum and electrovacuum with a $G_2$ group of symmetries
is contained in the work of Berger et al \cite{ber} (the condition of
timelike total EM is not explicited in \cite{ber}, but it follows from the
boundary conditions there imposed). It can be seen that if one has a
universe with timelike total EM, one needs to add some matter in order to make
the total EM spacelike \cite{car}, and thus, this ``quite general'' 
property that CTCs need spacelike total EM gives a precise notion of how much
energy it is needed for causality violation. A similar result in $3+1$ would,
of course, be of the greatest interest. 

\section{Some final comments}

The argument that we gave as supporting the property of the polarized surfaces
as ``finders'' of Cauchy Horizons is essentially the original one of Kim and
Thorne. Though it is widely used and it is usually  expected to hold under
very general conditions, up to our knowledge there is no rigorous proof of
it. Some parts of the analysis of the previous section are implicit in  
Cutler's work, so we now make contact with it. Cutler takes a point $p$
through which a SNG passes and chooses two charts of inertial like coordinates
(one chart for each particle). He then explicitly calculates the map $k^i
\rightarrow k^f$ as a non linear map $g(\phi)$ from the circle of null
directions at $p$ to itself, and uses the fact that $g$ has two fixed points
to obtain the tangent to the horizon (one fixed point corresponds to the
tangent to the future horizon, and the other one to the past horizon) and
reconstruct it using some symmetries of the spacetime. That is, his map
$g$ corresponds, essentially, to our map ${\cal L}$. We have here taken
advantage of the fact that ${\cal L}$ is linear, to see under which conditions
there are fixed null directions (the fixed points of $g$ correspond to the null
eigenvectors of ${\cal L}$); and we have  noted that ${\cal L}$
defines the total EM, so that the two fixed points that Cutler finds do not
depend on the details of the geometry of Gott's spacetime, but rather on the
property that its total EM is spacelike.

\section*{Acknowledgements}

The author thanks Sean Carroll for kindly reading
the manuscript and for his valuable comments. He also acknowledges A.
Ni, C. Valmont; and CONICET for financial support. This work was supported in 
part by grants from the National University of C\'ordoba, and from CONICOR,
and CONICET (Argentina).

\end{document}